\documentclass[aps,pra,twocolumn,floatfix]{revtex4}
\usepackage{bm,dcolumn,amsmath,graphicx}
\begin{document}
\title{Atomic properties of Cd-like and Sn-like ions for the development of frequency standards and search for the variation of the fine-structure constant}
\author{M. S. Safronova$^{1,2}$}
\author{V. A. Dzuba$^{3}$}
\author{V. V. Flambaum$^{3}$}
\author{U. I. Safronova$^{4,5}$}
\author{S. G. Porsev$^{1,6}$}
\author{M. G. Kozlov$^{6,7}$}

\affiliation {$^1$University of Delaware, Newark, Delaware, USA}
\affiliation {$^2$Joint Quantum Institute, NIST and the University of Maryland, College Park, Maryland, USA}
\affiliation {$^3$The University of New South Wales, Sydney, Australia}
\affiliation {$^4$University of Nevada, Reno, Nevada, USA}
\affiliation {$^5$University of Notre Dame, Notre Dame, Indiana, USA}
\affiliation {$^6$Petersburg Nuclear Physics Institute, Gatchina,  Russia}
\affiliation {$^7$St.\ Petersburg Electrotechnical University ``LETI'', St.\ Petersburg, Russia}
\date{\today}

\begin{abstract}
A high-precision relativistic calculations of  Cd-like Nd$^{12+}$, Sm$^{14+}$ and Sn-like Pr$^{9+}$, Nd$^{10+}$ atomic properties is
carried out using an approach that combines configuration interaction and a linearized coupled-cluster method. These ions
have long-lived metastable states with transitions accessible by laser excitations, relatively simple electronic structure,
high sensitivity to $\alpha$ variation, and stable isotopes. Breit and  QED corrections were included into the calculations.
Energies, transition wavelengths, electric- and magnetic-multipole reduced matrix elements, lifetimes, and sensitivity coefficients
$q$ and $K$ to the variation of the fine-structure constant $\alpha$ were obtained. A detailed study of uncertainties was performed.
Energies for  similar Cd-like Ba$^{8+}$, La$^{9+}$, Ce$^{10+}$, Pr$^{11+}$ and Sn-like Ba$^{6+}$ ions were calculated and  compared
with experiment for further tests of the accuracy.
\end{abstract}

\maketitle

\section{Introduction}
Last five years marked extraordinary improvements in both the accuracy and
stability of optical frequency standards~\cite{1,2,3}. The most accurate
trapped-ion clock based on quantum logic spectroscopy of an Al$^+$ ion was
demonstrated in 2010 \cite{1}. The fractional frequency uncertainty of
$8.6\times10^{-18}$ was reported. The optical frequency standard based on
$^{88}$Sr$^+$  trapped ion with the  total fractional frequency uncertainty of
$2.3\times10^{-17}$ was reported in \cite{DubMadZho13}. In 2013, Yb lattice
clock with instability of $8.6\times10^{-18}$ after only 7 hours of averaging
\cite{2} was reported. The $6.4\times10^{-18}$ accuracy was achieved with the
Sr optical lattice clock \cite{3} which represents a factor of 22  improvement
in comparison with the best previous optical lattice clock. Cryogenic Sr
optical lattice clocks with a relative frequency difference of $10^{-18}$ was
demonstrated in \cite{katori}.

Further development of even more precise frequency standards is essential for
new tests of fundamental physics, search for the variation of fundamental
constants, and very-long-baseline interferometry for telescope array
synchronization. The most precise laboratory test of variation of the
fine-structure constant $\alpha$ has been carried out by measuring the
frequency ratio of Al$^+$ and Hg$^+$ optical atomic clocks with a fractional
uncertainty of $5.2\times10^{-17}$ \cite{RosHumSch08}. Furthermore, more
precise clocks will enable the development of extremely sensitive
quantum-based tools for geodesy, hydrology, climate change studies,
inertial navigation, and tracking of deep-space probes \cite{2,3}.

This remarkable progress poses the question of what are the novel schemes for
the clock development that may achieve the accuracy at the next decimal point,
$10^{-19}$. %To the best of our knowledge, there are
We can single out two types of new clock
scheme proposals at the present time. The first set of proposals are for the
development of a nuclear clock ~\cite{peik2003} based on the $^{229}$Th
nuclear transition that has an unusually low first excitation energy of only
several eV making it accessible with laser excitation.  The second set of
proposals involves various transitions in highly-charged ions (HCI)
\cite{BerDzuFla10,DerDzuFla12,DzuDerFla12}. The estimates of potential
accuracy of clocks based on highly-charged ions and the $^{229}$Th  nuclear
transition are similar, but most HCI clock proposals do not have a complication of dealing
with the radioactive isotope.

Recent studies of uncertainties \cite{DerDzuFla12,DzuDerFla12} have shown that
the fractional uncertainty of the transition frequency in the clocks based on
HCIs can be smaller than 10$^{-19}$. Estimated sensitivity to the variation of
$\alpha$ for highly-charged ions approaches 10$^{-20}$ per year
\cite{DzuDerFla12}, which may allow for tests of spatial variation of the
fine-structure constant that may be indicated by the observational studies
\cite{WebKinMur11}.

While HCIs lack strong electric-dipole transitions for laser cooling, some
have strong M1 transitions. Moreover, sympathetic cooling may be employed
similar to the scheme used in Al$^+$ clock, which is cooled using laser-cooled
Be$^{+}$ or Mg$^+$ ions~\cite{1}. The experimental investigations toward the
sympathetic cooling of HCIs and the precision laser spectroscopy of forbidden
transitions are in progress
\cite{HobSolSuh11,AndCazNor13,SchVerWin12,VerSchWin13}. A cooling scheme
combining laser cooling of Be$^{+}$ ions and sympathetic cooling of Xe$^{44+}$
by Coulomb collisions with the cold Be$^+$ ions has been demonstrated in
\cite{GruHolSte01}. In 2011, the evaporative cooling of Ar$^{16+}$ in a
Penning trap was demonstrated~\cite{HobSolSuh11}. A novel extraction technique
based on the excitation of a coherent axial oscillation which allowed to
monitor the cooling process and to extract HCI bunches of high density and low
momentum spread was also demonstrated ~\cite{HobSolSuh11}. Laser cooling of
Mg$^+$ ions in a Penning trap for sympathetic cooling of highly-charged ions
was demonstrated in ~\cite{AndCazNor13}. Storage and cooling of highly-charged
ions require ultra-high vacuum levels. These can be obtained by cryogenic
methods, and  a linear Paul trap operating at 4~K capable of very long ion
storage times of about 30~h was recently developed in ~\cite{SchVerWin12,VerSchWin13}.
Capture and isolation of highly-charged ions in a unitary Penning trap
extracted from an electron beam ion trap (EBIT) at NIST was demonstrated in
~\cite{BreGuiTan13}. The  observed energy distribution was $~60$ times smaller
than typically expected for ions inside an EBIT without applying any active
cooling~\cite{BreGuiTan13}.

In a recent  work, we  proposed 10 highly-charged ions that belong to Ag-like,
In-like, Cd-like, and Sn-like isoelectronic sequences as candidates for the
development of next generation atomic clocks, search for variation of
fine-structure constant, and quantum information \cite{SafDzuFla14}. Ag-like
and In-like highly-charged ions have been further considered in
Ref.~\cite{AgIn}.

In this work, we carried out detailed high-precision study of Cd-like
Nd$^{12+}$, Sm$^{14+}$ and Sn-like Pr$^{9+}$, Nd$^{10+}$ highly-charged ions
using an approach that combines configuration interaction (CI) and a variant of the
coupled-cluster method. Breit and quantum electrodynamic (QED) corrections were included into the
calculations.   Our
calculations include energies, transition wavelengths, electric-dipole,
electric-quadrupole, electric-octupole, magnetic-dipole, magnetic-quadrupole,
magnetic-octupole reduced matrix elements, lifetimes, and sensitivity
coefficients to $\alpha$-variation $q$  and $K$. We carried out extensive
study of the uncertainties of our results. Two types of calculations were
carried out for Sn-like ions, treating these ions as systems with two and four
valence electrons to ensure that all important configurations were taken into
account. Energies for  similar Cd-like Ba$^{8+}$, La$^{9+}$, Ce$^{10+}$,
Pr$^{11+}$ and Sn-like Ba$^{6+}$ ions, where the experimental values are
available, were calculated and compared with experiment for further tests of
accuracy. Our values are in excellent agreement with experimental energies
from \cite{nist-web,cd-like-la-02,cd-like-ce-01} for Cd-like Ba$^{8+}$,
La$^{9+}$, Ce$^{10+}$, with similar level of agreement for all three ions.
However, we find a significant  discrepancy with experimental values from
\cite{cd-like-pr-03,cd-like-nd-05} for Pr$^{11+}$ and Nd$^{+12}$ which might
indicate a problem with the experimental level identification. Detailed study
of higher-order, Breit, QED, and higher partial wave contributions was carried
out to evaluate uncertainties of the final results for each ion.

We start with the brief description of the CI+all-order method used in this
work in Section~\ref{method}.   The results for Cd-like and Sn-like ions are
presented in Sections ~\ref{cd} and \ref{sn}, respectively.

\section{Method}
\label{method} The Cd-like ions are divalent systems with two valence
electrons above the $1s^2 2s^2 2p^6 3s^2 3p^6 3d^{10} 4s^2 4p^6 4d^{10}$ core.
We use a CI+all-order method developed in \cite{Koz04,SafKozJoh09} that
combines the modified linearized single-double coupled-cluster approach with
configuration interaction. The CI many-electron wave function is obtained
as a linear combination of all distinct states of a given angular momentum $J$
and parity~\cite{KotTup87}: %\cite{DzuFlaKoz96b}:
\begin{equation}
\Psi_{J} = \sum_{i} c_{i} \Phi_i\,.
\end{equation}
The CI + many-body perturbation theory (MBPT) approach developed in
~\cite{DzuFlaKoz96b} allows one to incorporate core excitations in the CI
method by including  perturbation theory terms into an effective Hamiltonian
$H^{\text{eff}}$. The one-body part $H_1$ is modified to include the
correlation potential $\Sigma_1$ that accounts for one-body part of the
core-valence correlations:
\begin{equation}\label{H1eff}
H_1 \rightarrow H_1+\Sigma_1
\end{equation} and
the two-body Coulomb interaction term $H_{2}$ is modified by including
the two-body part of core-valence interaction that represents
screening of the Coulomb interaction by valence electrons;
\begin{equation}\label{H2eff}
H_2 \rightarrow H_2+\Sigma_2.
\end{equation}
Then, the energies and wave functions of low-lying states are determined by
diagonalizing the effective Hamiltonian:
\begin{equation}
H^{\text{eff}}=H_{1} + H_2, \label{ham}
\end{equation}
where $H_{1}$ and $H_2$ are modified according to Eqs.\ \eqref{H1eff} and \eqref{H2eff}.
%represents the one-body and two-body parts of the Hamiltonian, respectively.
The matrix elements and other properties, such as polarizabilities, can be
determined using the resulting wave functions \cite{KotTup87}.

In the CI + all-order approach, the corrections to the effective Hamiltonian
$\Sigma_1$ and $\Sigma_2$ are calculated using a modified version of the
linearized coupled-cluster all-order method which allows to include dominant
core and core-valence correlation corrections to the effective Hamiltonian to
all orders and improve accuracy in comparison with the CI+MBPT method. The
detailed description of the CI+all-order method and all formulas are given in
~\cite{SafKozJoh09}.

When the CI space includes only two or three electrons, it can be made
essentially complete. For four-electron systems, we have developed an
efficient algorithm to construct a sufficiently complete set of
configurations. The CI+all-order method yielded accurate wave functions for
calculations of such atomic properties as lifetimes, polarizabilities,
hyperfine structure constants, etc, for a number of divalent and threvalent systems
\cite{SafKozJoh09,SafKozCla11,PorSafKoz12,SafPorKoz12,SafKozCla12,SafKozSaf12,SafPorCla12,SafSafPor13}.
The  spectra of the superheavy elements No, Lr and Rf with two, three, and
four valence electrons were recently presented by Dzuba {\it et al.\/}
\cite{DzuSafSaf14}.

We included the Breit interaction on the same footing as the Coulomb
interaction at the stage of constructing the basis set, and incorporated the Gaunt part of the Breit
interaction in the CI.  The QED radiative corrections to the energy levels are
included using the method described in \cite{FlaGin05}. We find the QED
contribution to be significant only for the configurations that contain
valence  $5s$ state, and omit it for Sn-like ions where none of the low-lying
configurations  contain $5s$ valence electron. The partial waves with $l_{max}=6$
are included in all summations in many-body perturbation theory or
coupled-cluster terms. Extrapolation of the $l>6$ contribution is carried out
following the method described in Ref.~\cite{AgIn}.

The lifetime of a state $a$ is calculated as
 \begin{align}\label{tau}
 \tau_a=\frac{1}{\sum_{b}
 A_{ab}},
 \end{align}
where the multipole transition rates $A_{ab}$ are related to the line
strengths $S_{ab}$. Explicit expressions are given in Ref.~\cite{AgIn}. In the
sum \eqref{tau} we account for the electric ($Ek$) and magnetic ($Mk$) transitions of the ranks $k= 1-3$.
%determined using
%\begin{eqnarray}
% A(E1) &=& \frac{2.02613\times 10^{18}} {(2J_a+1)\lambda ^{3}} \, \, S(E1),\\
%  A(M1) &=& \frac {2.69735\times 10^{13}}{(2J_a+1)\lambda ^{3}} \, \, S(M1),\\
%   A(E2)& =& \frac{1.11995\times 10^{18}}{(2J_a+1)\lambda ^{5}} \, \, S(E2),\\
%  A(M2)& =& \frac{1.49097 \times 10^{13}}{(2J_a+1)\lambda ^{5}} \, \, S(M2),\\
%   A(E3)& = &\frac{3.14441 \times 10^{17}}{(2J_a+1)\lambda ^{7}} \, \, S(E3),\\
%  A(M3) &= &\frac{4.18610 \times 10^{12}}{(2J_a+1)\lambda ^{7}} \, \, S(M3),
% \end{eqnarray}
%where the wavelength $\lambda$ is in \AA~and the line strength $S$ is in atomic units.

The sensitivity of the atomic transition frequency $\omega$ to the variation
of the fine-structure constant $\alpha$ can be quantified using a coefficient
$q$ defined as $
 \omega(x)=\omega_0+qx,
$
  where
$
 x \equiv \left(\frac{\alpha}{\alpha_0}\right)^2-1
$
and the frequency $\omega_0$ corresponds to the value of the fine-structure constant $\alpha_0$ at some initial point in time. It is convenient to also define dimensionless  enhancement factor $K=2q/\omega$. We follow the same procedure to calculate $q$
as in Ref.~\cite{AgIn}. Briefly, we carry out three calculations with different values of $\alpha$ for every ion considered in this work.
 In the first calculation, current CODATA value of $\alpha$ ~\cite{MohTayNew11} is used. In the other two calculations, the value
  of $\alpha^2$ is varied by $\pm$1\%. The value of $q$ is then determined as a numerical derivative.

\begin{table*}
\caption{\label{tab-cd-like} Energies of Cd-like Ba$^{8+}$, Nd$^{12+}$, and
Sm$^{14+}$ ions relative to their ground states evaluated using  the CI+all-order
method (in cm$^{-1}$). Contributions from higher-order Coulomb correlation (difference of the CI+all-order and CI+MBPT calculation),
estimated contributions of  higher partial waves ($l>6$),
the Breit and QED corrections are given separately in columns HO, Extrap, Breit, and QED.
Experimental results are from \cite{nist-web} for Ba$^{8+}$ and \cite{cd-like-nd-05} for Nd$^{12+}$.
Difference with experiment is given in cm$^{-1}$ and \% in columns ``Diff.'' Estimated absolute uncertainties of
theoretical calculations are given in columns ``Unc''. Theoretical and experimental wavelengths for
transitions to the ground states are given in last two columns in nm.}
\begin{ruledtabular}
\begin{tabular}{lrrrrrrrrrrrcc}
\multicolumn{1}{c}{Ion}&
\multicolumn{1}{c}{Level}&
\multicolumn{1}{c}{Expt}&
\multicolumn{1}{c}{CI+MBPT}&
\multicolumn{1}{c}{HO}&
\multicolumn{1}{c}{Extrap}&
\multicolumn{1}{c}{Breit}&
\multicolumn{1}{c}{QED}&
\multicolumn{1}{c}{Final}&
\multicolumn{1}{c}{Unc}& \multicolumn{1}{c}{Diff.}&
\multicolumn{1}{c}{Diff.\%}& \multicolumn{1}{c}{$\lambda_\textrm{th}$}&
 \multicolumn{1}{c}{$\lambda_\textrm{expt}$}\\
 \hline      \\[-0.4pc]
Ba$^{8+}$  &  $5s^2  \ ^1S_0$ &   0  &0     &0   & 0   &    0  &      0 &0     &  &       &      &     &       \\
           &  $5s5p  \ ^3P_0$ &116992&119350&-1434& 1   &    349&   -506 &117760&  & -768   &-0.66\% &84.92&   85.48 \\
           &  $5s5p  \ ^3P_1$ &122812&124971&-1289& 3   &   298 &   -500 &123483&  & -671   &-0.55\% &80.98&   81.43 \\
           &  $5s5p  \ ^3P_2$ &142812&145610&-1532& 13  &   25  &   -461 &143655&  & -843   &-0.59\% &69.61&   70.02 \\
           &  $5s5p  \ ^1P_1$ &175712&175440& 646 & 21  &    33 &    -466&175674&  & 38     &0.02\%  &56.92&   56.91 \\
           &  $4f5s  \ ^3F_2$ &237170&238470& 809 & -898&   -912&   -558 &236911&  &  259   &0.11\%  &42.21&   42.16  \\
           &  $4f5s  \ ^3F_3$ &237691&239062& 792 & -896&   -971&   -555 &237432&  &  259   &0.11\%  &42.12&   42.07  \\
           &  $4f5s  \ ^3F_4$ &238547&240038& 755 & -891&  -1078&  -548  &238276&  &  271   &0.11\%  &41.97&   41.92  \\
           &  $4f5s  \ ^1F_3$ &245192&246989& 768 & -900&  -1033&  -566  &245258&  &  -66   &-0.03\% &40.77&   40.78   \\  [0.5pc]
Nd$^{12+}$ &  $5s^2 \ ^1S_0$  &     0&0  &  0&    0&       0&     0 &       0&      &     &       &                   \\
           &  $5s4f \ ^3F_2$  & 77162&81730  &1258& -1128&  -1407&   -983&   79469&   600&     2307&  2.99\% &125.8(9)& 129.6\\
           &  $5s4f \ ^3F_3$  & 78443&83119  &1246& -1127&  -1489&   -979&   80769&   610&     2326&  2.97\% &123.8(9)& 127.5\\
           &  $5s4f \ ^3F_4$  & 81440&86393  &1198& -1121&  -1763&   -978&   83730&   650&     2290&  2.81\% &119.4(9)& 122.8\\
           &  $5s4f \ ^1F_3$  & 87312&92519  &1218& -1128&  -1699&   -959&   89951&   640&     2639&  3.02\% &111.2(8)& 114.5\\
           &  $5s5p \ ^3P_0$  &156417&161505  &-1511&    -7&    650&   -970&  159667&  1020&     3250&  2.08\%&62.6(4)   & 63.9\\
           &  $5s5p \ ^3P_1$  &165482&170161  &-1223&    -5&    584&   -970&  168547&   800&     3065&  1.85\%&59.3(3)   & 60.4\\
           &  $5s5p \ ^3P_2$  &204685&210480  &-1620&     8&     77&   -970&  207976&  1400&     3291&  1.61\%&48.1(3)   & 48.9\\
           &  $5s5p \ ^1P_1$  &245748&245644  &821&    16&     83&   -970&  245594&   320&     -154& -0.06\% &40.7(1)   & 40.7\\    [0.5pc]
 Sm$^{14+}$ & $4f^2 \  ^3H_4$  & &0 &    0  &    0  &   0  &   0     &     0  &    &&&         & \\
            &  $5s4f \ ^3F_2$  & &611 &-2437  &1140   &1658  &1201     &2172    &850  &&&4600(1300)&\\
            &  $5s4f \ ^3F_3$  & &2355 &-2447  &1140   &1574  &1204     &3826    &840 &&&2614(470) & \\
            &  $4f^2 \ ^3H_5$  & &5378 &  -61  & 4     &-408  &26       &4939    &100 &&&2025(40)  & \\
            &  $5s4f \ ^3F_4$  & &7392 &-2498  &1146   &1190  &1233     &8463    &810 &&&1182(100) & \\
            &  $4f^2 \ ^3F_2$  & &9320 & -194  &40     &35    &6        &9207    &50  &&&1086(6)  & \\
            &  $4f^2 \ ^3H_6$  & &10797 & -124  &9      &-827  &51       &9906    &210&&&1010(20) & \\
            &  $4f^2 \ ^3F_3$  & &12974 & -235  &39     &-274  &27       &12532   &90 &&&798(6)   &  \\
            &  $4f^2 \ ^1G_4$  & &13620 & -223  &36     &-352  &26       &13108   &110&&&763(6)   & \\
            & $5s4f \  ^1F_3$  & &13207 &-2479  &1141  &1254   &1214    &14337   &810 &&&698(40)  & \\
            & $4f^2 \  ^3F_4$  & &20633 & -299  &43    &-715   &56      &19717   &200 &&&507(5)   & \\
 \end{tabular}
\end{ruledtabular}
\end{table*}

\begin{table*}
\caption{\label{tab-cd-like1} Comparison of theoretical energies with
experiment for  Cd-like Ba$^{8+}$ \cite{nist-web}, La$^{9+}$
\cite{cd-like-la-02}, Ce$^{10+}$ \cite{cd-like-ce-01}, Pr$^{11+}$
\cite{cd-like-pr-03},  and Nd$^{12+}$ \cite{cd-like-nd-05} ions relative to the $5s^2\,\,^1\!S_0$ ground state (in cm$^{-1}$).
Actual (in cm$^{-1}$) and relative (in \%) differences with experiment are given for all states. The states are listed in the same order
for all ions. Fine structure intervals for $5s5p$ and $4f5s$ triplets are listed in the last four rows. }
\begin{ruledtabular}
\begin{tabular}{lrrrrrrrrrrrrrrr}
\multicolumn{1}{c}{Level} &
\multicolumn{3}{c}{Ba$^{8+}$} &
\multicolumn{3}{c}{La$^{9+}$} &
\multicolumn{3}{c}{Ce$^{10+}$} &
\multicolumn{3}{c}{Pr$^{11+}$} &
\multicolumn{3}{c}{Nd$^{12+}$} \\
\multicolumn{1}{c}{} &
\multicolumn{1}{c}{Theory} & \multicolumn{1}{c}{Diff.} &\multicolumn{1}{c}{\%} &\multicolumn{1}{c}{Theory} & \multicolumn{1}{c}{Diff.} &\multicolumn{1}{c}{\%} &
\multicolumn{1}{c}{Theory} & \multicolumn{1}{c}{Diff.} &\multicolumn{1}{c}{\%} &\multicolumn{1}{c}{Theory} & \multicolumn{1}{c}{Diff.} &\multicolumn{1}{c}{\%} &
\multicolumn{1}{c}{Theory} & \multicolumn{1}{c}{Diff.} &\multicolumn{1}{c}{\%} \\
\hline \\ [-0.3pc]
$5s5p\,\,^3\!P_0$ & 117760 & -768  & -0.7 & 128226  & -811 & -0.6  & 138718 & -867  & -0.6  & 149252  & -2934 & -2.0 & 159667 & -3250 & -2.1\\

$5s5p\,\,^3\!P_1$ &  23483 & -671  & -0.6 & 134719  & -697 & -0.5  & 145996 & -729  & -0.5  & 157329  & -2774 & -1.8 & 168547 & -3065 & -1.9\\

$5s5p\,\,^3\!P_2$ & 143655 & -843  & -0.6 & 158956  & -885 & -0.6  & 173838 & -891  & -0.5  & 191188  & -3020 & -1.6 & 207976 & -3291 & -1.6\\

$5s5p\,\,^1\!P_1$ & 175674 &   38  &  0.0 & 192451  &   33 &  0.0  & 209702 &    1  &  0.0  & 227471  &   -24 & -0.0 & 245594 &   154 &  0.1\\

$4f5s\,\,^3\!F_2$ & 236911 &  259  &  0.1 & 207147  &  165 &  0.1  & 170661 &   41  &  0.0  & 127955  & -1969 & -1.6 &  79469 & -2307 & -3.0\\

$4f5s\,\,^3\!F_3$ & 237432 &  259  &  0.1 & 207865  &  159 &  0.1  & 171521 &   91  &  0.1  & 129067  & -1984 & -1.6 &  80769 & -2326 & -3.0\\

$4f5s\,\,^3\!F_4$ & 238276 &  271  &  0.1 & 209118  &  171 &  0.1  & 173364 &  104  &  0.1  & 131378  & -1964 & -1.5 &  83730 & -2290 & -2.8\\

$4f5s\,\,^1\!F_3$ & 245258 &  -66  & -0.0 & 215933  & -144 & -0.1  & 179860 & -232  & -0.1  & 137767  & -2297 & -1.7 &  89951 & -2639 & -3.0 \\

$^3\!P_1-\,^3\!P_0$ & 5723 &   97  &  1.7 &   6493  &  114 &  1.7  &   7278 &  138  &  1.9  &   8077  &   160 &  1.9 &   8880 &   185 &  2.0 \\
$^3\!P_2-\,^3\!P_1$ &20172 & -172  & -0.9 &  24237  & -188 & -0.8  &  28571 & -162  & -0.6  &  33859  &  -246 & -0.7 &  39429 &  -226 & -0.6 \\
$^3\!F_3-\,^3\!F_2$ &  521 &    0  &  0.0 &    718  &   -6 & -0.8  &    860 &   50  &  5.5  &   1112  &   -15 & -1.4 &   1300 &   -19 & -1.5 \\
$^3\!F_4-\,^3\!F_3$ &  844 &   12  &  1.4 &   1253  &   12 &  1.0  &   1752 &   13  &  0.7  &   2311  &    20 &  0.9 &   2961 &    36 &  1.2 \\
\end{tabular}
\end{ruledtabular}
\end{table*}

\begin{table}
\caption{\label{tab-cd-like-q} Transition energies $\omega$ and  sensitivity coefficients $q$
for Cd-like ions relative to the ground state evaluated using the CI+all-order
method in cm$^{-1}$; $K=2q/\omega$ is the enhancement factor.}
\begin{ruledtabular}
\begin{tabular}{lrrrrr}
\multicolumn{1}{c}{Ion}&
\multicolumn{1}{c}{Level}&
\multicolumn{1}{c}{$\omega$}&
\multicolumn{1}{c}{$q$}&
\multicolumn{1}{c}{$K$}\\
\hline   \\[-0.4pc]
Nd$^{12+}$  &   $5s^2 \ ^1S_0$  &   0   &   0   &       \\
    &   $5s4f \ ^3F_2$  &   79469   &   101461  &   2.6 \\
    &   $5s4f \ ^3F_3$  &   80769   &   102325  &   2.5 \\
    &   $5s4f \ ^3F_4$  &   83730   &   105340  &   2.5 \\
    &   $5s4f \ ^1F_3$  &   89951   &   105827  &   2.4 \\
    &   $5s5p \ ^3P_0$  &   159667  &   14175   &   0.2 \\
    &   $5s5p \ ^3P_1$  &   168547  &   19465   &   0.2 \\[0.5pc]
Sm$^{14+}$  &   $4f^2 \  ^3H_4$     &   0   &   0   &      \\
    &    $5s4f \ ^3F_2$     &   2172    &   -127720 &   -118    \\
    &    $5s4f \ ^3F_3$     &   3826    &   -126746 &   -66 \\
    &    $4f^2 \ ^3H_5$     &   4939    &   4917    &   2.0 \\
    &    $5s4f \ ^3F_4$     &   8463    &   -121952 &   -29 \\
    &    $4f^2 \ ^3F_2$     &   9207    &   1324    &   0.3 \\
    &    $4f^2 \ ^3H_6$     &   9906    &   9295    &   1.9 \\
    &    $4f^2 \ ^3F_3$     &   12532   &   4954    &   0.8 \\
    &    $4f^2 \ ^1G_4$     &   13108   &   4508    &   0.7 \\
    &   $5s4f \  ^1F_3$     &   14337   &   -121525 &   -17 \\
    &   $4f^2 \  ^3F_4$     &   19717   &   10045   &   1.0
\end{tabular}
\end{ruledtabular}
\end{table}

 \begin{table*}
\caption{\label{tab-life-nd12} The CI+all-order multipole reduced matrix elements $Z$ (in a.u.),
transition rates $A_r$ (in s$^{-1}$), and lifetimes $\tau$ (in sec) in Cd-like Nd$^{12+}$ and Sm$^{14+}$ ions.
Transition energies (in cm$^{-1}$) and corresponding wavelengths (in nm) are obtained from the theoretical energies given in  Table~\ref{tab-cd-like}. The numbers in brackets represent powers of 10.}
\begin{ruledtabular}
\begin{tabular}{lrrrcrrrrr}
\multicolumn{1}{c}{Ion}&
\multicolumn{1}{c}{Level}&
\multicolumn{2}{c}{Transition}&
\multicolumn{1}{c}{}&
\multicolumn{1}{c}{Energy}&
\multicolumn{1}{c}{$\lambda$}&
\multicolumn{1}{c}{$Z$}&
\multicolumn{1}{c}{$A_{r}$}&
\multicolumn{1}{c}{$\tau$}\\
\hline \\ [-0.5pc]
Nd$^{12+}$ & $5s4f\ ^3F_{2}$   &  $5s2  \ ^1S_{0}$ &  $5s4f\ ^3F_{2}$&  M2 &     79469&    125.8&      0.00012 &    1.312[-11] &    7.622[+10]\\[0.2pc]
            & $5s4f\ ^3F_{3}$   &  $5s4f \ ^3F_{2}$ &  $5s4f\ ^3F_{3}$&  M1 &      1300&     7692&      2.49401 &    5.266[-02] &    18.90 \\[0.2pc]
            & $5s4f\ ^3F_{4}$   &  $5s4f \ ^3F_{3}$ &  $5s4f\ ^3F_{4}$&  M1 &      2961&     3377&      2.50909 &    4.898[-01] &    2.042\\
            &                   &  $5s4f \ ^3F_{2}$ &  $5s4f\ ^3F_{4}$&  E2 &      4261&     2347&      0.07926 &    1.098[-07] &    \\[0.2pc]
            & $5s4f\ ^1F_{3}$   &  $5s4f \ ^3F_{2}$ &  $5s4f\ ^1F_{3}$&  M1 &     10482&    954.0&      0.61326 &    1.669[+00] &    0.410\\
            &                   &  $5s4f \ ^3F_{4}$ &  $5s4f\ ^1F_{3}$&  M1 &      6221&     1607&      0.61858 &    3.549[-01] &    \\[0.2pc]
            & $5s5p\ ^3P_{0}$   &  $5s4f \ ^3F_{2}$ &  $5s5p\ ^3P_{0}$&  E2 &     80198&    124.7&      0.58339 &    1.265[+02] &    7.905[-03]\\[0.2pc]
            & $5s5p\ ^3P_{1}$   &  $5s^2  \ ^1S_{0}$&  $5s5p\ ^3P_{1}$&  E1 &    168547&     59.3&      0.42601 &    5.868[+08] &    1.710[-09] \\[0.2pc]
            & $5s5p\ ^3P_{2}$   &  $5s4f \ ^3F_{4}$ &  $5s5p\ ^3P_{2}$&  E2 &    124246&     80.5&      1.14090 &    8.632[+02] &    6.054[-04]\\
            &                   &  $5s^2  \ ^1S_{0}$&  $5s5p\ ^3P_{2}$&  M2 &    207976&     48.1&      5.97200 &    4.138[+00] &    \\
            &                   &  $5s4f \ ^3F_{2}$ &  $5s5p\ ^3P_{2}$&  E2 &    128507&     77.8&      0.21718 &    3.703[+01] &    \\
            &                   &  $5s5p \ ^3P_{0}$ &  $5s5p\ ^3P_{2}$&  E2 &     48309&    207.0&      1.23160 &    8.940[+00] &    \\
            &                   &  $5s5p \ ^3P_{1}$ &  $5s5p\ ^3P_{2}$&  M1 &     39429&    253.6&      1.49436 &    7.384[+02] &    \\[0.2pc]
            &  $5s5p\ ^1P_{1}$  &  $5s2  \ ^1S_{0}$ &  $5s5p\ ^1P_{1}$&  E1 &    245594&     40.7&      1.39510 &    1.947[+10] &    5.136[-11]\\
            &                   &  $5s4f \ ^1F_{3}$ &  $5s5p\ ^1P_{1}$&  E2 &    155643&     64.2&      0.89639 &    2.740[+03] &    \\
            &                   &  $5s5p \ ^3P_{0}$ &  $5s5p\ ^1P_{1}$&  M1 &     85927&    116.4&      0.42382 &    1.024[+03] &    \\
            &                   &  $5s5p \ ^3P_{2}$ &  $5s5p\ ^1P_{1}$&  M1 &     37618&    265.8&      0.47796 &    1.093[+02] &    \\[0.4pc]\hline

 Sm$^{14+} $&$ 4f5s\ ^3F_{2}$  &$ 4f^2\ ^3H_{4}$&$   4f5s\ ^3F_{2}$& M2&    2172&   4604&    0.03516&   1.782[-14]&  5.613[+13]\\[0.2pc]
  &$ 4f5s\ ^3F_{3}$  &$ 4f^2\ ^3H_{4}$&$   4f5s\ ^3F_{3}$& E1&    3826&   2614&    0.00092&   1.366[-02]&   8.514  \\
  &                  &$4f5s \ ^3F_{2}$&$   4f5s\ ^3F_{3}$& M1&    1654&   6046&    2.43986&   1.038[-01]&         \\  [0.2pc]

  &$ 4f^2\ ^3H_{5}$  &$ 4f^2\ ^3H_{4}$&$   4f^2\ ^3H_{5}$& M1&    4939&   2025&    3.19913&   3.024[+00]&   0.331 \\ [0.2pc]
  &$ 4f5s\ ^3F_{4}$  &$4f5s \ ^3F_{3}$&$   4f5s\ ^3F_{4}$& M1&    4637&   2157&    2.45429&   1.800[+00]&   0.556 \\ [0.2pc]

  &$ 4f^2\ ^3F_{2}$  &$4f5s \ ^3F_{3}$&$   4f^2\ ^3F_{2}$& E1&    5381&   1858&    0.00652&   2.682[+00]&   0.373 \\
  &                  &$4f^2 \ ^3H_{4}$&$   4f^2\ ^3F_{2}$& E2&    9207&   1086&    0.49808&   3.676[-04]&         \\ [0.2pc]

  &$ 4f^2\ ^3H_{6}$  &$ 4f^2\ ^3H_{5}$&$   4f^2\ ^3H_{6}$& M1&    4967&   2013&    3.26426&   2.709[+00]&   0.369 \\ [0.2pc]

  &$ 4f^2\ ^3F_{3}$  &$4f5s \ ^3F_{2}$&$   4f^2\ ^3F_{3}$& E1&   10360&    965&    0.00205&   1.351[+00]&   0.328 \\
  &                  &$4f5s \ ^3F_{4}$&$   4f^2\ ^3F_{3}$& E1&    4069&   2458&    0.00601&   7.041[-01]&         \\
  &                  &$ 4f^2\ ^3F_{2}$&$   4f^2\ ^3F_{3}$& M1&    3325&   3007&    2.52457&   9.028[-01]&         \\
  &                  &$ 4f^2\ ^3H_{4}$&$   4f^2\ ^3F_{3}$& M1&   12532&    798&    0.10761&   8.782[-02]&         \\ [0.2pc]

  &$ 4f^2\ ^1G_{4}$  &$4f5s \ ^3F_{3}$&$   4f^2\ ^1G_{4}$& E1&    9282&   1077&    0.00373&   2.506[+00]&   0.338 \\
  &                  &$ 4f^2\ ^3H_{5}$&$   4f^2\ ^1G_{4}$& M1&    8169&   1224&    0.52656&   4.530[-01]&         \\
  &                  &$ 4f^2\ ^3F_{3}$&$   4f^2\ ^1G_{4}$& M1&     576&  17361&    1.95991&   2.200[-03]&         \\ [0.2pc]

  &$ 4f5s\ ^1F_{3}$  &$ 4f^2\ ^3H_{4}$&$   4f5s\ ^1F_{3}$& E1&   14337&    698&    0.00475&   1.929[+01]&   0.0410\\
  &                  &$4f5s \ ^3F_{2}$&$   4f5s\ ^1F_{3}$& M1&   12165&    822&    0.80183&   4.460[+00]&         \\
  &                  &$4f5s \ ^3F_{4}$&$   4f5s\ ^1F_{3}$& M1&    5874&   1702&    0.80877&   5.108[-01]&         \\
  &                  &$ 4f^2\ ^3F_{4}$&$   4f5s\ ^1F_{3}$& E1&    1229&   8137&    0.01529&   1.255[-01]&         \\ [0.2pc]

  &$ 4f^2\ ^3F_{4}$  &$4f5s \ ^1F_{3}$&$   4f^2\ ^3F_{4}$& E1&    5380&   1859&    0.01635&   9.374[+00]&  0.0648\\
  &                  &$4f5s \ ^3F_{3}$&$   4f^2\ ^3F_{4}$& E1&   15891&    629&    0.00120&   1.309[+00]&         \\
  &                  &$ 4f^2\ ^3H_{5}$&$   4f^2\ ^3F_{4}$& M1&   14778&    677&    0.40844&   1.614[+00]&         \\
  &                  &$ 4f^2\ ^3F_{3}$&$   4f^2\ ^3F_{4}$& M1&    7185&   1392&    1.67986&   3.137[+00]&         \\ [0.2pc]
\end{tabular}
\end{ruledtabular}
\end{table*}

\section{Cd-like ions}
\label{cd}
The $5s-4f$ level crossing in Cd-like ions happens for Nd$^{12+}$ - Sm$^{14+}$ ions. The order of levels in previous ions of the
Cd-like isoelectronic sequence, such as  Ba$^{8+}$, is $5s^2$, $5s5p$, and $5s4f$. It changes to $5s^2$, $5s4f$, and  $5s5p$ for
Nd$^{12+}$. The $4f^2$ becomes the ground state for Sm$^{14+}$, with other low-lying levels belonging to either  $4f^2$ or $5s4f$
configurations. In order to evaluate the uncertainties of our values, we carried out several calculations which allowed us to separate the effect of higher orders, Breit interaction, contributions of higher partial waves, and QED. The contribution of the
higher orders is evaluated as the difference of the CI+all-order and CI+MBPT results. The Breit and QED contributions are
calculated as the difference of the results with and without the inclusion of these effects. The contribution of the higher $(l>6)$
partial waves (labeled ``Extrap'') is estimated to be equal to the contribution of the $l=6$ partial wave following our empiric rule obtained for Ag-like ions (see \cite{AgIn} for a detailed discussion of the extrapolation). The contribution of the $l=6$ partial wave is obtained as the difference of two calculations where all intermediate sums in the all-order and MBPT terms are restricted to $l_{\rm max}=6$  and $l_{\rm max}=5$. The resulting four contributions are listed separately in Table~\ref{tab-cd-like}. The final theoretical results are listed in ``Final'' column.

We develop several  methods to estimate the accuracy of our calculations.
First, we assume that the uncertainty of each of the four corrections (HO, Extrap, Breit, and QED) does not exceed 25\%, and add 25\%
of each correction in quadrature to estimate the total uncertainty. In Ag-like and In-like  ions, such estimates are significantly larger (by a factor 2-4) than our actual difference with the experiment for all three ions listed in Table~I of \cite{SafDzuFla14}. For Ba$^{+8}$ $5s4f$ states, which are of  most interest for the present work,
 such estimate gives about 400~cm$^{-1}$, while our differences with experiment are 70-270~cm$^{-1}$. Therefore, we can expect that
such procedure will give reasonable estimates of uncertainties for the $5s4f$ states of Nd$^{12+}$.

 \begin{table*}
\caption{\label{tab-sn-like}
Comparison of CI+all-order energies of Sn-like Ba$^{6+}$,  Pr$^{9+}$, and
Nd$^{10+}$ ions relative to the ground state calculated as two-valence-electron (2-valence) and four-valence-electron
(4-valence) systems (in cm$^{-1}$).
In the two-valence electron calculation, the $5s$ shell is taken to be a core shell.
Contributions from higher-order Coulomb correlation (difference of the CI+all-order and CI+MBPT calculations),
estimated contributions of  higher partial waves ($l>6$),  and Breit corrections are given separately in columns labeled
``HO'', ``Extrap'', and ``Breit''. Differences between ``4-valence'' and ``2-valence'' final values are given in last column.}
\begin{ruledtabular}
\begin{tabular}{crlrrrrrrrrrrr}
\multicolumn{1}{c}{}&
\multicolumn{2}{c}{}&
\multicolumn{5}{c}{2-valence calculation}&
  \multicolumn{5}{c}{4-valence calculation}&  \multicolumn{1}{c}{Diff.}\\
 \multicolumn{1}{c}{Ion}&
 \multicolumn{1}{r}{Term}&
 \multicolumn{1}{l}{$J$}&
 \multicolumn{1}{c}{CI+MBPT}&
\multicolumn{1}{c}{HO}&
\multicolumn{1}{c}{Extrap}&
\multicolumn{1}{c}{Breit}&
\multicolumn{1}{c}{Final}&
  \multicolumn{1}{c}{CI+MBPT}&
\multicolumn{1}{c}{HO}&
\multicolumn{1}{c}{Extrap}&
\multicolumn{1}{c}{Breit}&
\multicolumn{1}{c}{Final}   &
\multicolumn{1}{c}{}     \\
\hline   \\[-0.4pc]
Ba$^{6+}$  & $5p^2$& $^3P_0$&0    &     0&    0 &      0&     0 &    0& 0    &   0 &    0  &  0   &0     \\
           & $5p^2$& $^3P_1$&15554&  153 &    13&  -242 &  15477&15704& -103 &   10 & -238 &15372 &-105  \\
           & $5p^2$& $^3P_2$&21228&  187 &    13&  -267 &  21161&21788& -108 &   9  & -268 &21422 &261 \\
           & $5p^2$& $^1D_2$&42400&  -26 &    24&  -510 &  41888&43143& -233 &   17 & -507 &42420 &532 \\
           & $5p^2$& $^1S_0$&62976& -1466&    25&  -505 &  61030&62342& -260 &   21 & -501 &61602 &572  \\   [0.5pc]
Pr$^{9+}$  &$5p^2$&$^3P_0$      &0    & 0&     0&      0&      0&0& 0    &    0 &   0   &  0     & 0     \\
           &$5p4f$&$^3G_3$      &20050&  2994& -1078&  -1750&  20216&21865 & 2810 & -1032& -1748 &  21895 &  1679 \\
           &$5p4f$&$^3F_2$      &22664&  2489&  -862&  -1519&  22772&24172 & 2291 & -829 & -1435 &  24199 &  1427  \\
           &$5p4f$&$^3F_3$      &25607&  2844& -1072&  -2017&  25362&27233 & 2804 & -1026& -2009 &  27002 &  1640  \\
           &$5p4f$&$^3F_4$      &27727&  2943& -1080&  -2054&  27536&29622 & 2801 & -1033& -2048 &  29343 &  1806  \\
           &$5p^2$&$^3P_1$      &28712&  193 &    16&   -409&  28512&28962 & -135 &   14 & -405  &  28436 &  -76    \\
           &$5p^2$&$^3P_2$      &35831&  856 &  -252&   -782&  35653&36697 & 615  & -243 & -852  &  36217 &  564   \\
           &$5p4f$&$^1F_3$      &54104&  2728& -1066&  -2179&  53588&55735 & 2680 & -1023& -2172 &  55220 &  1632  \\  [0.5pc]
 Nd$^{10+}$&$4f^2     $&$ ^3H_4$ &0   & 0&      0&      0&      0&0&     0&     0  &     0 &     0  &    0      \\
           &$5p4f     $&$ ^5G_3$ &2605& -4037&  1076&   1920&   1564&1953& -4305&   1025 &  1887 &   560  &   -1004   \\
           &$4f^2     $&$ ^3H_5$ &3432& -80  &     1&   -294&   3059&3405& -58  &      2 &  -291 &   3058 &   -1       \\
           &$5p4f     $&$ ^1D_2$ &5975& -3629&  1009&   1704&   5060&5171& -3823&    923 &  1768 &   4040 &   -1020    \\
           &$4f^2     $&$ ^3H_6$ &6982& -167 &     6&   -599&   6222&6930& -128 &      8 &  -590 &   6219 &   -3       \\
           &$5p4f     $&$ ^3F_3$ &8448& -3861&  1037&   1471&   7095&7853& -4222&   1063 &  1687 &   6382 &   -713      \\
           &$4f^2+5p4f$&$ ^3F_2$ &8263& -694 &   113&    231&   7914&8323& -594 &    144 &  134  &   8007 &   93       \\
           &$4f^2+5p4f$&$ ^5G_4$ &9391& -3048&   908&   1102&   8353&8845& -3330&    801 &  1309 &   7624 &   -729     \\
 \end{tabular}
\end{ruledtabular}
\end{table*}

 \begin{table}
\caption{\label{tab-sn-like2}
Comparison of the CI+all-order energies of Sn-like Ba$^{6+}$ relative to the ground state calculated as two-valence-electron (2-val)
and four-valence-electron (4-val) system (in cm$^{-1}$). Experimental results~\cite{nist-web} are listed in column labeled ``Expt.''.
The columns $\Delta_{\rm{2val}}$ and  $\Delta_{\rm{4val}}$ give differences between 2-val and 4-val calculations and experiment.
In the two-valence-electron calculation, the $5s$ shell is taken to be a core shell.
The column labeled ``Ave.'' gives average of the 4-val and 2-val calculations.
In last column the difference of averaged results with experiment is presented.}
\begin{ruledtabular}
\begin{tabular}{rrrrrrrr}
\multicolumn{1}{c}{Level}&
\multicolumn{1}{c}{2-val}&
\multicolumn{1}{c}{4-val}&
\multicolumn{1}{c}{Expt.}&
\multicolumn{1}{c}{$\Delta_{\textrm{2val}}$}&
\multicolumn{1}{c}{$\Delta_{\textrm{4val}}$}&
\multicolumn{1}{c}{Ave.}&
\multicolumn{1}{c}{$\Delta_{\textrm{Ave}}$}\\
% \multicolumn{4}{c}{}&
% \multicolumn{1}{c}{}&
% \multicolumn{1}{c}{2-val}&
% \multicolumn{1}{c}{4-val}&
% \multicolumn{1}{c}{}&
% \multicolumn{1}{c}{Ave.}\\
 \hline      \\[-0.4pc]
 $5p^2\,\,^3\!P_0$ &      0&       0 &       0 &   0 &   0 &     0  &    0    \\
 $5p^2\,\,^3\!P_1$ &  15477&   15372 &   15507 &  30 & 135 &  15425 &    82   \\
 $5p^2\,\,^3\!P_2$ &  21161&   21422 &   21499 &  338&  77 &  21291 &   208   \\
 $5p^2\,\,^1\!D_2$ &  41888&   42420 &   42514 &  626&  94 &  42154 &   360   \\
 $5p^2\,\,^1\!S_0$ &  61030&   61602 &   61083 &  53 & -519&  61316 &  -233   \\
 \end{tabular}
\end{ruledtabular}
\end{table}

\begin{table*}
\caption{\label{tab-sn-like3} Comparison of the CI+all-order energies of Sn-like Pr$^{9+}$ and Nd$^{10+}$ relative to the ground
state calculated as two-valence-electron (2-val) and four-valence-electron (4-val) system (in cm$^{-1}$). The final numbers, which are the
average of two calculations are listed in the column labeled ``Final''. Estimated absolute uncertainties of
the respective values are given in columns ``Unc'' in cm$^{-1}$. Wavelengths for transitions to the ground state
and their uncertainties (in parenthesis) are given in last column in nm.}
\begin{ruledtabular}
\begin{tabular}{crlrrrrrrcccc}
\multicolumn{1}{c}{Ion}&
\multicolumn{1}{c}{Term}&
 \multicolumn{1}{c}{$J$}&
\multicolumn{1}{c}{2-val}&
  \multicolumn{1}{c}{Unc.}&
\multicolumn{1}{c}{4-val}&
  \multicolumn{1}{c}{Unc.}&
%\multicolumn{1}{c}{Ave.}&
 \multicolumn{1}{c}{Final}&
   \multicolumn{1}{c}{Unc.}&
\multicolumn{1}{c}{$\lambda$}\\
 \hline      \\[-0.4pc]
Pr$^{9+}$ &$5p^2$&$^3P_0$&      0&     &       0&    0&     0 &    0 &      \\
          &$5p4f$&$^3G_3$&  20216&  540&  21895 &  450& 21055 &  840& 475(18) \\
          &$5p4f$&$^3F_2$&  22772&  430&  24199 &  370& 23485 &  710& 426(13) \\
          &$5p4f$&$^3F_3$&  25362&  580&  27002 &  570& 26182 &  820& 382(12) \\
          &$5p4f$&$^3F_4$&  27536&  560&  29343 &  590& 28440 &  900& 352(11) \\
          &$5p^2$&$^3P_1$&  28512&  160&  28436 &  320& 28474 &  320& 351.2(5) \\
          &$5p^2$&$^3P_2$&  35653&  240&  36217 &  380& 35935 &  380& 278(2) \\
          &$5p4f$&$^1F_3$&  53588&  710&  55220 &  710& 54404 &  820& 184(3) \\[0.5pc]
 Nd$^{10+}$&$4f^2     $&$ ^3H_4$&     0&    0&  0     &    0 &  0   &   0&                \\
           &$5p4f     $&$ ^5G_3$& 1564&  1100&  560   & 1300 &1062  &1300&                \\
           &$4f^2     $&$ ^3H_5$& 3059&   220&  3058  &  210 &3059  &220 &  3270(220)\\
           &$5p4f     $&$ ^1D_2$& 5060&   980&  4040  &  1100&4550  &1100&  2200(430) \\
           &$4f^2     $&$ ^3H_6$& 6222&   460&  6219  &  430 &6221  &460 &  1610(110) \\
           &$5p4f     $&$ ^3F_3$& 7095&  1200&  6382  &  1320&6738  &1320&  1480(240) \\
           &$4f^2+5p4f$&$ ^3F_2$& 7914&   270&  8007  &  240 &7960  &270 &  1260(40) \\
           &$4f^2+5p4f$&$ ^5G_4$& 8353&   940&  7624  &  1070&7989  &1070&  1250(150) \\
 \end{tabular}
\end{ruledtabular}
\end{table*}

\begin{table}
\caption{\label{tab-sn-like-q} Transition energies, $\omega$, and  sensitivity coefficients $q$
for Sn-like ions relative to the ground state evaluated in the CI+all-order
approximation in cm$^{-1}$; $K=2q/\omega$ is the enhancement factor.}
\begin{ruledtabular}
\begin{tabular}{lrlrrr}
\multicolumn{1}{c}{Ion}&
\multicolumn{1}{c}{Term}&\multicolumn{1}{c}{$J$}&
\multicolumn{1}{c}{$\omega$}&
\multicolumn{1}{c}{$q$}&
\multicolumn{1}{c}{$K$}\\
\hline   \\[-0.4pc]
Pr$^{9+}$&  $5p^2$&$^3P_0$&        0   &   0   &       \\
         &  $5p4f$&$^3G_3$&        20216   &   42721   &   4.2 \\
         &  $5p4f$&$^3F_2$&        22772   &   42865   &   3.8 \\
         &  $5p4f$&$^3F_3$&        25362   &   47076   &   3.7 \\
         &  $5p4f$&$^3F_4$&        27536   &   37197   &   2.7 \\
         &  $5p^2$&$^3P_1$&        28512   &   47483   &   3.3 \\[0.5pc]
Nd$^{10+}$& $4f^2     $&$ ^3H_4$&   0      &           &       \\
         &  $5p4f     $&$ ^5G_3$&  1564    &   -81052  &   -104    \\
         &  $4f^2     $&$ ^3H_5$&  3059    &   3113    &   2.0 \\
         &  $5p4f     $&$ ^1D_2$&  5060    &   -60350  &   -24 \\
         &  $4f^2     $&$ ^3H_6$&  6222    &   5930    &   1.9 \\
         &  $5p4f     $&$ ^3F_3$&  7095    &   -63285  &   -18 \\
         &  $4f^2+5p4f$&$ ^3F_2$&  7914    &   -17809  &   -4.5    \\
         &  $4f^2+5p4f$&$ ^5G_4$&  8353    &   -39672  &   -9.5    \\
    \end{tabular}
\end{ruledtabular}
\end{table}

\begin{table}
\caption{\label{tab-mult-2-4}
Absolute values of multipole reduced matrix elements obtained by  CI+all-order two-valence-electrons (2-val) and
four-valence-electrons (4-val) calculations in Sn-like Pr$^{9+}$ ion (in a.u.).}
\begin{ruledtabular}
\begin{tabular}{lrrcc}
\multicolumn{1}{l}{}&
\multicolumn{2}{c}{Transition}&
\multicolumn{1}{c}{2-val}&
\multicolumn{1}{c}{4-val}\\
\hline \\ [-0.4pc]
M1& $5p^2\ ^3P_{0}$&$   5p^2\ ^3P_{1}$&      1.273 &   1.274 \\
M1& $5p4f\ ^3F_{2}$&$   5p^2\ ^3P_{1}$&      0.323 &   0.359 \\
M1& $5p4f\ ^3G_{3}$&$   5p4f\ ^3F_{2}$&      0.320 &   0.312 \\
E2& $5p2 \ ^3P_{0}$&$   5p2 \ ^3P_{2}$&      1.901 &   1.864 \\
E2& $5p^2\ ^3P_{0}$&$   5p4f\ ^3F_{2}$&      0.232 &   0.110 \\
E2& $5p^2\ ^3P_{1}$&$   5p4f\ ^1F_{3}$&      0.623 &   0.577 \\
E2& $5p4f\ ^3G_{3}$&$   5p^2\ ^3P_{1}$&      0.183 &   0.170 \\
E2& $5p4f\ ^3F_{3}$&$   5p^2\ ^3P_{1}$&      1.133 &   1.071 \\
E2& $5p4f\ ^3G_{3}$&$   5p4f\ ^3F_{3}$&      0.117 &   0.089 \\
E2& $5p4f\ ^3F_{2}$&$   5p4f\ ^3F_{4}$&      0.662 &   0.686 \\
E2& $5p4f\ ^3G_{3}$&$   5p4f\ ^1F_{3}$&      2.109 &   2.071 \\
M3& $5p^2\ ^3P_{1}$&$   5p^2\ ^3P_{2}$&     13.369 &  14.162 \\
M3& $5p^2\ ^3P_{0}$&$   5p4f\ ^3G_{3}$&      0.427 &   0.328 \\
M3& $5p^2\ ^3P_{0}$&$   5p4f\ ^3F_{3}$&      4.906 &   5.246 \\
\end{tabular}
\end{ruledtabular}
\end{table}

\begin{table*}
\caption{\label{tab-life-pr9} The CI+all-order  multipole matrix elements $Z$ (in a.u.), transition rates $A_r$ (in s$^{-1}$),
and lifetimes $\tau$ (in sec) in Sn-like Pr$^{9+}$ and Nd$^{10+}$ ions. Transition energies $\Delta E$ (in cm$^{-1}$) and wavelengths
$\lambda$ (in nm) are obtained from final energy values given by Table~\ref{tab-sn-like3}. The numbers in brackets represent powers of 10.}
\begin{ruledtabular}
\begin{tabular}{crrrrrrrrr}
\multicolumn{1}{c}{Ion}&
\multicolumn{1}{r}{Term}&
\multicolumn{2}{c}{Transition}&
\multicolumn{1}{c}{}&
\multicolumn{1}{c}{$\Delta E$}&
\multicolumn{1}{c}{$\lambda$}&
\multicolumn{1}{c}{$Z$}&
\multicolumn{1}{c}{$A_{r}$}&
\multicolumn{1}{c}{$\tau$}\\
\hline \\[-0.4pc]
Pr$^{9+}$&$5p4f\ ^3G_{3}$& $5p^2\ ^3P_{0}$&  $5p4f\ ^3G_{3}$& M3&      21055&     474.9&     0.42712&   2.001[-15]&   4.997[+14] \\ [0.2pc]

 &$5p4f\ ^3F_{2}$& $5p^2\ ^3P_{0}$&  $5p4f\ ^3F_{2}$& E2&      23485&     425.8&     0.23230&   8.635[-03]&   51.8      \\
 &               & $5p4f\ ^3G_{3}$&  $5p4f\ ^3F_{2}$& M1&       2430&     4115.&     0.31981&   7.917[-03]&             \\ [0.2pc]

 &$5p4f\ ^3F_{3}$& $5p4f\ ^3F_{2}$&  $5p4f\ ^3F_{3}$& M1&       2697&     3708.&     1.67447&   2.120[-01]&   4.718     \\ [0.2pc]

 &$5p4f\ ^3F_{4}$& $5p4f\ ^3G_{3}$&  $5p4f\ ^3F_{4}$& M1&       7385&     1354.&     1.89256&   4.324[+00]&   0.227     \\
 &               & $5p4f\ ^3F_{3}$&  $5p4f\ ^3F_{4}$& M1&       2258&     4429.&     1.46389&   7.394[-02]&             \\ [0.2pc]

 &$5p^2\ ^3P_{1}$& $5p^2\ ^3P_{0}$&  $5p^2\ ^3P_{1}$& M1&      28474&     351.2&     1.27262&   3.362[+02]&   2.975[-3] \\ [0.2pc]

 &$5p^2\ ^3P_{2}$& $5p^2\ ^3P_{0}$&  $5p^2\ ^3P_{2}$& E2&      35935&     278.3&     1.90140&   4.852[+00]&  0.0838     \\
 &               & $5p4f\ ^3G_{3}$&  $5p^2\ ^3P_{2}$& M1&      14880&     672.0&     0.20488&   7.461[-01]&             \\
 &               & $5p4f\ ^3F_{2}$&  $5p^2\ ^3P_{2}$& E2&      12450&     803.2&     0.60441&   2.448[-03]&             \\
 &               & $5p4f\ ^3F_{3}$&  $5p^2\ ^3P_{2}$& M1&       9753&     1025.&     0.90901&   4.135[+00]&             \\
 &               & $5p^2\ ^3P_{1}$&  $5p^2\ ^3P_{2}$& M1&       7461&     1340.&     0.99055&   2.198[+00]&             \\ [0.2pc]

 &$5p4f\ ^1F_{3}$& $5p4f\ ^3F_{2}$&  $5p4f\ ^1F_{3}$& M1&      30919&     323.4&     1.02292&   1.192[+02]&   7.382[-3] \\
 &               & $5p4f\ ^3G_{3}$&  $5p4f\ ^1F_{3}$& E2&      33349&     299.9&     2.10930&   2.936[+00]&             \\
 &               & $5p4f\ ^3F_{4}$&  $5p4f\ ^1F_{3}$& M1&      25964&     385.1&     0.25958&   4.545[+00]&             \\
 &               & $5p^2\ ^3P_{2}$&  $5p4f\ ^1F_{3}$& M1&      18469&     541.4&     0.60127&   8.776[+00]&             \\ [0.2pc]
\hline
Nd$^{10+}$& $5p4f     \ ^5G_{3}$&   $4f^2\  ^3H_{4}$&  $5p4f     \ ^5G_{3}$   &M1&   1062&    9416&    0.06522&   1.963[-05]&  5.094[+04]\\[0.2pc]
          & $4f^2     \ ^3H_{5}$&   $4f^2\  ^3H_{4}$&  $4f^2     \ ^3H_{5}$   &M1&   3059&    3269&    3.21593&   7.259[-01]&   1.378    \\[0.2pc]
          & $5p4f     \ ^1D_{2}$&   $5p4f\  ^5G_{3}$&  $5p4f     \ ^1D_{2}$   &M1&   3488&    2867&    0.41514&   3.945[-02]&   25.35    \\
          &                     &   $4f^2\  ^3H_{4}$&  $5p4f     \ ^1D_{2}$   &E2&   4550&    2198&    0.66642&   1.940[-05]&            \\[0.2pc]
          & $4f^2     \ ^3H_{6}$&   $4f^2\  ^3H_{5}$&  $4f^2     \ ^3H_{6}$   &M1&   3162&    3163&    3.26836&   7.007[-01]&   1.427    \\[0.2pc]
          & $5p4f     \ ^3F_{3}$&   $5p4f\  ^1D_{2}$&  $5p4f     \ ^3F_{3}$   &M1&   2188&    4570&    1.94864&   1.533[-01]&   3.916    \\
          &                     &   $5p4f\  ^5G_{3}$&  $5p4f     \ ^3F_{3}$   &M1&   5676&    1762&    0.36934&   9.612[-02]&            \\
          &                     &   $4f^2\  ^3H_{4}$&  $5p4f     \ ^3F_{3}$   &M1&   6738&    1484&    0.07115&   5.967[-03]&            \\[0.2pc]
          & $4f^2+5p4f\ ^3F_{2}$&   $5p4f\  ^1D_{2}$&  $4f^2+5p4f\ ^3F_{2}$   &M1&   3410&    2933&    0.29565&   1.870[-02]&   43.92    \\
          &                     &   $5p4f\  ^5G_{3}$&  $4f^2+5p4f\ ^3F_{2}$   &M1&   6898&    1450&    0.04406&   3.437[-03]&            \\
          &                     &   $4f^2\  ^3H_{4}$&  $4f^2+5p4f\ ^3F_{2}$   &E2&   7960&    1256&    0.93855&   6.305[-04]&            \\[0.2pc]
          & $4f^2+5p4f\ ^5G_{4}$&   $5p4f\  ^5G_{3}$&  $4f^2+5p4f\ ^5G_{4}$   &M1&   6927&    1444&    1.57004&   2.456[+00]&   0.365    \\
          &                     &   $4f^2\  ^3H_{4}$&  $4f^2+5p4f\ ^5G_{4}$   &M1&   7989&    1252&    0.38552&   2.271[-01]&            \\
          &                     &   $4f^2\  ^3H_{5}$&  $4f^2+5p4f\ ^5G_{4}$   &M1&   4930&    2028&    0.36808&   4.865[-02]&            \\
          &                     &   $5p4f\  ^3F_{3}$&  $4f^2+5p4f\ ^5G_{4}$   &M1&   1251&    7994&    1.39879&   1.148[-02]&            \\[0.2pc]
\end{tabular}
\end{ruledtabular}
\end{table*}

In the second approach of evaluating the uncertainties, we use  the reference ion, Ba$^{8+}$, to estimate the uncertainties in the calculations for the other ion. We estimate the uncertainty as the sum of the following: (1) difference of the theoretical and experimental energies for the reference ion and (2) difference in the sum of all four corrections between the reference and the current ion.
For the $5s5p$ states of Nd$^{12+}$,  we use this second approach to estimate the uncertainties and also find that these estimates are significantly smaller than our difference with the experiment.

The agreement of the $5s4f$ energies with the experiment  \cite{nist-web} for Ba$^{8+}$ is excellent and is of the same relative magnitude (0.1\%) as in the case of $4f$ states of Ag-like Ba$^{9+}$ ion. The $5s5p$ energies agree  with experiment to about 0.6\%.
However, the differences with the experiment for Nd$^{12+}$ \cite{cd-like-nd-05} energies are anomalously large, 1.6-3\% for all states listed in Table~\ref{tab-cd-like} except $5s5p\,\,^1\!P_1$ which is in excellent agreement with experiment. These differences are much larger than our estimated upper bound on the uncertainty of our results listed in column  ``Unc''.

To explore the discrepancy of our energies with experiment for Nd$^{12+}$, we calculated the energies of the other three ions of  Cd-like isoelectronic sequence, La$^{9+}$,  Ce$^{10+}$, and Pr$^{11+}$, and compared the results with the experimental values from \cite{cd-like-la-02,cd-like-ce-01,cd-like-pr-03}. We present the comparison of theoretical energies with experiment for all 5 consecutive ions of Cd-like isoelectronic sequence,  Ba$^{8+}$ \cite{nist-web}, La$^{9+}$ \cite{cd-like-la-02}, Ce$^{10+}$ \cite{cd-like-ce-01}, Pr$^{11+}$ \cite{cd-like-pr-03},  and
Nd$^{12+}$ \cite{cd-like-nd-05} in Table~\ref{tab-cd-like1}. All energies are given relative to the $5s^2\,^1\!S_0$ ground state in cm$^{-1}$. Actual (in cm$^{-1}$) and relative (in \%) differences with experiment are given for all states. The states are listed in the same order for all 5 ions for the convenience of presentation. The actual order of states starts to change for Ce$^{10+}$.
The fine structure splittings of the $5s5p$ and $4f5s$ triplets are listed in the last four rows. Table~\ref{tab-cd-like1} clearly illustrates the abrupt shift in the agreement with experiment between the first three and last two ions. It appears that all levels with the exception of the $5s5p\,^1\!P_1$ suddenly shift by about 2000~cm$^{-1}$ for Pr$^{11+}$. We note that the $5s^2\,\,^1\!S_0-5s5p\,\,^1\!P_1$ is the only strong easily identifiable line from all of the states considered here. The identification of other numerous ultraviolet (UV) lines
is a very difficult task carried out in \cite{cd-like-la-02,cd-like-ce-01,cd-like-pr-03} using the Cowan code. It may be possible that change in the order of levels for Pr$^{11+}$ resulted in some identification problem. Since our calculations are carried out in the same way for all ions, we find abrupt 2000~cm$^{-1}$ shift in accuracy to be unlikely. Further measurements are needed to resolve this problem.
We use the first (25\%) approach to estimate the accuracy of our calculations for Sm$^{14+}$ energies since $4f^2$ configuration is not present among measured low-lying levels of Ba$^{8+}$, and this reference ion cannot be used for Sm$^{14+}$.

 The CI+all-order sensitivity coefficients $q$ for Cd-like ions obtained as described in Section~\ref{method}
are given in Table~\ref{tab-cd-like-q}. All energy and $q$ values are given relative to the ground state in cm$^{-1}$.
The CI+all-order energies and $q$ coefficients are used to calculate  enhancement factors $K=2q/\omega$ given in the last row of the table.
The enhancement factors are very large for all  transitions from the $5s4f$ levels to the ground state for Sm$^{14+}$ due to large $q$ and small transition energies.
The calculation of $q$ for Ag-like ions \cite{AgIn} demonstrated that the effect of correlation is small for the cases where $q$ are large, i.e. all cases of interest.
Therefore, the uncertainties in large values of $K$ will be dominated by the uncertainties in the transition energies, in particular where they exceed 2-3\%. Then, the relative uncertainty in $K$ for $4f^2\,^3\!H_4 - 5s4f$ transitions can be estimated as the relative uncertainty in the corresponding transition energy.
 We note that $q$ values are positive for Nd$^{12+}$  and negative for Sm$^{14+}$.  This creates additional enhancements for $\alpha$ variation search if the relative transition frequencies in Nd$^{12+}$/Sm$^{14+}$ are monitored.

The CI+all-order  multipole reduced matrix elements $Z$, transition rates $A_r$, and lifetimes $\tau$ in Cd-like Nd$^{12+}$ and Sm$^{14+}$ ions are
given in Table~\ref{tab-life-nd12}.  We use
theoretical energies in transition rate and lifetime calculations. The numbers in brackets represent powers of 10.
The strongest transition from the first excited levels of both ions is $M2$, leading to the extremely long lifetimes. The case of Nd$^{12+}$ is very similar to Ag-like Nd$^{13+}$ discussed in Ref.~\cite{AgIn} but the wavelengths are further in UV.
Next excited states in both ions live also very long, with 20~s and 8.5~s lifetimes for Nd$^{12+}$ and Sm$^{14+}$, respectively.

\section{Sn-like ions}
\label{sn} The Sn-like ions, considered in this work, may be treated either as
divalent systems with  $1s^2 2s^2 2p^6 3s^2 3p^6 3d^{10} 4s^2 4p^6 4d^{10} 5s^2$ core
or systems with four valence electrons (then, the $5s$ electrons are in the
valence field). We carry out both calculations to ensure that all dominant
configurations are taken into account. We refer to the results of these
calculations as 2-val and  4-val, respectively. We carried out the same
calculations for the Ba$^{6+}$ ion, which is the last ion in Sn isoelectronic
sequence where experimental energies are available. Unfortunately, the
experimental data for this ion are limited to the fine-structure of the $5p^2$
configuration. The results of 2-val and 4-val calculations for Sn-like ions
are summarized in Table~\ref{tab-sn-like} where we list the  energies of Sn-like Ba$^{6+}$, Pr$^{9+}$, and
Nd$^{10+}$ ions relative to the ground state.
Contributions from higher-order Coulomb correlation (difference of the CI+all-order and CI+MBPT calculations),
estimated contributions of  higher partial waves ($l>6$),  and
Breit interaction are given separately in columns labeled ``HO'', ``Extrap'', and ``Breit''.
QED contribution is considered to be negligible for these states.

We find a technical complication in applying CI+all-order method to the Sn-like ions.
Both the CI+MBPT and CI+all-order methods are based on the Brilloiun-Wigner variant of the
MBPT, rather than the Rayleigh-Schr\"{o}dinger  to avoid  nonsymmetric effective Hamiltonian and the
problem of intruder states. In the Brilloiun-Wigner variant of
MBPT, the effective Hamiltonian is symmetric and accidentally small
denominators do not arise, but the many-body corrections to the Hamiltonian  $\Sigma_1$ and $\Sigma_2$
become energy dependent. Solving the equation for $H^{\rm eff}$ we are able to find these energies.
But since we use the single-particle perturbation theory, more simple and practical approach is to set this energy,
$\tilde\varepsilon$, to be the Dirac-Fock energy of the lowest orbital for the particular partial wave
(see Refs.~\cite{SafKozJoh09,DzuFlaKoz96b} for more details).

For all Cd-like calculations carried out in this work, this approach works perfectly fine.
However, we find that the use of the lowest $5d$ energies, as $\tilde\epsilon$,
is not appropriate for the 2-val calculations for the Sn-like ions. This is because the $5s$ state, which is treated
as a core state in 2-val calculations, has small excitation energy, very close to the excitation energy of the $5d$ state.
This leads to   extremely small energy denominator in the expression for $\hat \Sigma$, e.~g.,
$\varepsilon_{5s} + \varepsilon_{5d} - \varepsilon_{5p} - \varepsilon_{5p} \simeq 0$. This means that the sum of two
single-electron energies $\varepsilon_{5s} + \varepsilon_{5d}$ is a poor approximation for
the two-electron energies of low states of Sn-like ions which must enter the expression for $\hat \Sigma$.
Therefore, we use the $5p_{1/2}$ energies for $\tilde\varepsilon$ in the expressions for $\hat \Sigma$ operator for $ns$ and $nd$
states.
% However, we  find that the use of the lowest $5d$ energies as
%  $\tilde\epsilon_{i}$ leads to failure in the construction of the
%  effective Hamiltonian already at the MBPT level. Therefore, we use
%  $5p_{1/2}$ energies for $\tilde\epsilon_{i}$ of the $ns$ and $nd$
%  states.
The $4f_j$ energies are used for the $nf$ states.

We have tested the sufficient completeness of the four-electron configuration space by carrying  out three
calculations with increasing number of configurations. The first run contained only double excitations into the valence space
from a few main configurations. % of interest.
Two larger runs were selectively constructed by allowing extra excitations into the valence space from several hundreds most important configurations. Thus, triple and quadrupole excitations from initial configurations were effectively included. The differences between the results  of first two runs were less then 350 cm$^{-1}$. The differences between second and final largest runs with 23000 and 32000 configurations, respectively, were less than 10~cm$^{-1}$ indicating sufficient saturation of the configuration space.

We provide a detailed comparison of 2-val and 4-val results with the experiment for Ba$^{6+}$~\cite{nist-web} in Table~\ref{tab-sn-like2}.
 While we find a good agreement of both results with experiment, it is unclear  whether 2-val or 4-val calculations provide better accuracy.
The agreement with experiment is different for the four states.
% The differences in agreement between the four states are
Most likely this is caused by the admixture of configurations that cannot be described as divalent $5s^2nln'l'$ states.
It appears that the average of both calculations produces the results that are the most consistent with experiment for all states.
Unfortunately, we have no comparison with experiment  for the $5p4f$ and  $4f^2$ configurations which are of interest for the present work.

The ions of interest in the Sn-like isoelectronic sequence are Pr$^{9+}$ and Nd$^{10+}$ where the $5p^2$ and $5p4f$ or $5p4f$ and $4f^2$
levels become very close due to the $5p-4f$ level crossing. The case of Pr$^{9+}$ is particularly interesting, since the several lowest
metastable levels have transitions to the ground state in the optical range. The ground and first excited states of Nd$^{10+}$
are extremely close and the resulting uncertainty is on the order of the transition energy. While our calculations place $4f^2$ to be
the ground state, the higher-order corrections are particularly large in this case, almost 3 times that of the transition energy,
which might lead to the placement of the $5p4f$ $J=3$ as the ground state.

Determination of the uncertainties is difficult for these ions due to complete lack of data for comparison. We also observe strong
cancellations between various large corrections. Therefore, adding 25\% of all corrections in quadrature may significantly overestimate the uncertainties. We take the average of the 25\% estimate and the total sum of all corrections as an uncertainty for all levels and list these values.
The uncertainties are independently evaluated for 2-val and 4-val calculations following this prescription.
Comparison of the CI+all-order energies of Sn-like Pr$^{9+}$ and
Nd$^{10+}$ relative to the ground state calculated as two-valence-electron and four-valence-electron system is given in
Table~\ref{tab-sn-like3}. The final numbers, which are the average of two calculations are listed in the column ``Final''.
Estimated uncertainties of all values are given in columns ``Unc''.  Wavelengths for transitions to the ground state are given in last columns in nm.
The sensitivity coefficients $q$ for Sn-like Pr$^{9+}$ and Nd$^{10+}$ ions are given in Table~\ref{tab-sn-like-q} together with the corresponding CI+all-order transition energies and $K$ enhancement factors.

 Comparison of multipole matrix elements
obtained from  CI+all-order 2-val and 4-val
calculations in Sn-like  Pr$^{9+}$ ion is given in Table~\ref{tab-mult-2-4}.
The differences are few percent for most transitions, but significant
for weak transitions, such as $E2$ $5p^2\,\,^3\!P_0 - 5p4f\,\,^3\!F_2$.
It is expected since weak transitions are more sensitive to the admixtures of configurations which are omitted in the 2-val calculation.

The CI+all-order  multipole matrix elements $Z$, transition rates $A_r$, and lifetimes $\tau$  in Sn-like Pr$^{9+}$ and Nd$^{10+}$ions
are presented in Table~\ref{tab-life-pr9}. Energies are final results presented in Table~\ref{tab-sn-like3}.
The numbers in brackets represent powers of 10.  We highlight the case of Pr$^{9+}$ where the lowest metastable state, $5p4f\,\,^3\!G_3$,
has a very long lifetime with $M3$ 495~nm transition to the ground state being in the optical range. Next two levels, $5p4f\,\,^3\!F_2$
and $5p4f\,\,^3\!F_3$, also have optical transitions to the ground state and are metastable  with 59~s and 5.3~s lifetimes.
A relatively strong $M1$ transition to the ground state from $5p^2\,\,^3\!P_1$ level at 351~nm may be potentially used for
cooling and probing. Our Nd$^{10+}$ transition property calculations assume that $4f^2~J=4$ level is the ground state.
While several low levels of Nd$^{10+}$ are long-lived, the corresponding transitions are all far in the infrared.

\section{Conclusion}

We carried out detailed high-precision study of  Cd-like Nd$^{12+}$, Sm$^{14+}$ and Sn-like Pr$^{9+}$, Nd$^{10+}$ atomic properties using a hybrid approach that combines configuration interaction and a linearized coupled-cluster method. These highly-charged ions are of interest for future  experimental studies aimed at the development of ultra-precise atomic clocks and search for $\alpha$ variation.
Energies, transition wavelengths, electric- and magnetic-multipole reduced matrix elements, lifetimes, and the sensitivity coefficients to
$\alpha$ variation, $q$ and $K$, were calculated. Several methods to evaluate uncertainties of the results were developed.

\section*{Acknowledgement}
We thank C. W. Clark, C. Monroe, J. Tan, Yu. Ralchenko, J. R. Crespo L\'{o}pez-Urrutia and P. Beiersdorfer for useful discussions. M.S.S. thanks School of Physics at the University of New South Wales, Sydney, Australia for hospitality and acknowledges support from Gordon Godfrey Fellowship, UNSW. This work was supported in part by US NSF Grant No.\ PHY-1212442. M.G.K. acknowledges support from RFBR Grant No.\ 14-02-00241.
The work was partly supported by the Australian Research Council.

%\bibliography{bibfile2013}
 \end{document}